\documentclass[
nofootinbib,
 amsmath,amssymb,
 aps,
pra,
twocolumn
]{revtex4-1}

\usepackage{dcolumn}
\usepackage{bm}
\usepackage{epsfig}
\usepackage{graphicx}
\usepackage{amsfonts}
\usepackage[figuresright]{rotating}
\usepackage{psfrag}
\usepackage{float} 
\usepackage{footmisc}

\input epsf.sty

\newcommand{\be}{\begin{equation}}
\newcommand{\ee}{\end{equation}}
\newcommand{\ba}{\begin{array}}
\newcommand{\ea}{\end{array}}
\newcommand{\bqa}{\begin{eqnarray}}
\newcommand{\eqa}{\end{eqnarray}}

\begin{document}

\title{Controlling the flow of light using inhomogeneous effective gauge field that emerges from dynamic modulation}
\author{Kejie Fang}
\altaffiliation[Current address: ]{Thomas J. Watson, Sr., Laboratory of Applied Physics, California Institute of Technology, Pasadena, CA 91125}
\homepage{https://sites.google.com/site/bacbever/}
\affiliation{Department of Physics, Stanford University, Stanford,
California 94305, USA}
\author{Shanhui Fan}
\affiliation{Department of Electrical Engineering, Stanford
University, Stanford, California 94305, USA}

\begin{abstract}
We show that the effective gauge field for photons provides a versatile platform for controlling the flow of light. As an example we consider a photonic resonator lattice where the coupling strength between nearest neighbor resonators are harmonically modulated. By choosing different spatial distributions of the modulation phases, and hence imposing different inhomogeneous effective magnetic field configurations, we numerically demonstrate a wide variety of propagation effects including negative refraction, one-way mirror, and on and off-axis focusing.  Since the effective gauge field is imposed dynamically after a structure is constructed, our work points to the importance of the temporal degree of freedom for controlling the spatial flow of light. 

\end{abstract}

\maketitle

It was recently recognized that when a photonic structure undergoes dynamic refractive index modulation, the phase of the modulation creates an effective gauge potential and effective magnetic field for photons \cite{our1,our2,our3}. The effective magnetic field can induce photonic phenomena similar to charged particles under real magnetic field, such as a photonic one-way edge mode \cite{our2} and a photonic de Haas-van Alphen effect \cite{our4}. In this paper, we further show that the use of \emph{inhomogeneous} effective gauge fields provide additional degrees of freedom in controlling the flow of light. As examples, we show that one can achieve negative refraction, one-way mirrors, circulators, and focusing, based on the same resonator lattice structure subject to different configurations of inhomogeneous effective gauge fields.  

Tailoring the propagation of light has been a central goal of nano-photonic research, which is critical for applications in on-chip communications and information processing \cite{pcbook}. Examples of previous studies include the use of waveguide arrays \cite{silberberg,lieven}, photonic crystals \cite{pc1,pc2,pc3,pc4} and meta-materials \cite{meta1,meta2,meta3} to achieve various beam propagation effects within these structures. Moreover, by introducing inhomogeneity into these structures \cite{ho1,ho2}, one can realize photon flow that emulates electron motion under electric field \cite{blochosc, zener, kivshar}.

Complementary to these works, which have largely focused on spatial degrees of freedom, our results here show that temporal degrees of freedom in a dynamic structure can also be quite useful in the control of electromagnetic wave propagations in space. Unlike the spatial (i.e. the structural) degrees of freedom, which are mostly defined by fabrication processes, the modulation phases can be readily changed in the dynamic structure, after the structure is constructed. Moreover, non-reciprocity, or time-reversal symmetry breaking, which is difficult to achieve in static structures unless magneto-optical materials are used, arises rather naturally in dynamically-modulated structures. Therefore, our approach of using photonic gauge field provides additional degrees of flexibility in controlling light propagation.

\begin{figure}[H]
\centering\epsfig{file=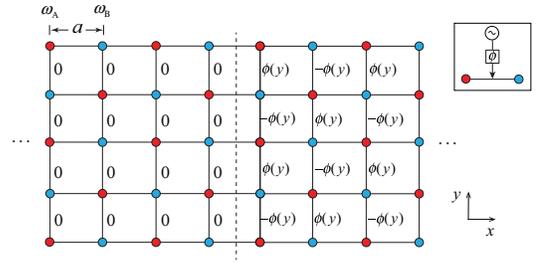,clip=1,width=0.8\linewidth,angle=0}\vspace{0.3cm}\hspace{0.3cm}
\caption{(Color online). A resonator lattice consisting of two kinds of resonators (red and blue dots) with dynamically modulated nearest neighbor coupling. The lattice is divided into two regions: the left region has an effective gauge potential $A_y=0$ and the right region has an effective gauge potential $A_y=\phi(y)/a$. Inset: the phase of the modulated coupling between nearest neighbor resonators corresponds to the RF wave that generates the modulation, and can be externally set by RF generators or RF phase shifters. }
\label{lattice}
\end{figure}

To illustrate the idea we use the model system introduced in \cite{our2} that consists of a two-dimensional photonic resonator lattice as shown in Fig. \ref{lattice}. The lattice has a square unit cell and each unit cell contains two resonators $A$ and $B$ with different resonant frequencies $\omega_A$ and $\omega_B$, respectively. We assume only nearest-neighbor coupling with a form of $V\textrm{cos}(\Omega t+\phi)$, where $V$ is the coupling strength, $\Omega$ and $\phi$ are the frequency and phase of the modulation. The Hamiltonian $H(t)$ of this resonator lattice is \bqa\label{Ht} H(t)&=&\omega_A\sum\limits_i a^\dagger_ia_i+\omega_B\sum\limits_j b^\dagger_jb_j\\\nonumber &&+\sum\limits_{\langle ij \rangle}V\textrm{cos}(\Omega t+\phi_{ij})(a^\dagger_ib_j+b^\dagger_ja_i), \eqa where $a_i^\dagger (a_i)$ and $b_j^\dagger (b_j)$ are the creation (annihilation) operators of the $A$ and $B$ resonators, respectively, and $\phi_{ij}$ is the phase of the modulation between resonators at nearest neighbor site $i$ and $j$.

We will assume an on-resonance modulation, i.e. $\Omega = |\omega_A - \omega_B|$, and stay in the regime $V\ll \Omega$ such that the rotation wave approximation applies. In such a case, when $\phi_{ij}=0$ everywhere, the structure is periodic and is described by a Floquet band structure \cite{our4}, $\epsilon(k_x,k_y) =V[\textrm{cos}(ak_x)+\textrm{cos}(ak_y)]$, where  $\epsilon$ is quasi-energy, $a$ is the separation between two nearest neighbor resonators, and $\vec k=k_x\hat{e}_x+k_y\hat{e}_y$ is the Bloch wavevector, which has no direct and simple relation to the free space wavelength or wavevector of light.

In this system as described by Eq. \eqref{Ht}, an effective gauge field arises from the spatial distribution of the modulation phase $\phi_{ij}$. This can be seen by going to a rotating frame, in which case the modulation phases then appear as the phases of the coupling constants in a time-independent tight-binding model, which gives rise to a gauge field structure through the Peierls substitution \cite{our2}. The value of the effective gauge field along the bond between sites $i$ and $j$ is determined by $\vec A=\vec l_{ij}\phi_{ij}/a$ \cite{our1}, where $\vec l_{ij}$ is a unit vector that points from site $i$ in the $A$ sub-lattice to site $j$ in $B$ sub-lattice. Furthermore, a non-uniform $\phi_{ij}$ distribution can create an effective magnetic field. The effective magnetic flux through a plaquette is defined as $B_\textrm{eff}=\oint \vec A\cdot d\vec l/a^2$ \cite{our2}, where the integration is along the sides of a plaquette. The consequences of having a uniform magnetic field in the lattice has been explored in Ref. \cite{our2,our4}. 

In this paper, we study the $\phi_{ij}$ distribution as shown in Fig. \ref{lattice}, which corresponds to a non-uniform magnetic field as we will see later. The lattice in Fig. \ref{lattice} can be separated into left and right regions. In the left region, the phases on the bonds along both $x-$ and $y-$axes are all zero. Therefore, the left region has zero effective gauge potential and zero effective magnetic field. We excite a beam by placing a source in the left region. In the right region, the phases on the bonds along the $x-$axis are zero, but the phases on the bonds along the $y-$axis is a function $\phi(y)$ of $y$ and alternate between positive and negative values. The different spatial configurations of $\phi(y)$ then correspond to different configurations of effective gauge potential and effective magnetic field in the right region. By choosing different modulation phase distribution in the right region, i.e. by choosing different $\phi(y)$, we can achieve versatile control of light propagation effects, for the beam incident from the left.

To simulate the motion of light in the presence of a source in this dynamically modulated resonator lattice we solve the coupled-mode equation \cite{our4}:
\be\label{sch} i\frac{d|\psi\rangle}{dt}=H(t)|\psi\rangle+|s\rangle,\ee 
where $|\psi\rangle=[\sum\limits_{i} v_i(t)a^\dagger_i+\sum\limits_{j}v_j(t)b^\dagger_j]|0\rangle$ is the photon state with $v_{i(j)}(t)$ being the amplitude at site $i(j)$. In our simulations, we use a source $|s\rangle$ that takes the form: 
\begin{widetext}
\be\label{source} |s\rangle=\sum\limits_{x,y}e^{-((x-x_0)^2+(y-y_0)^2)/w^2}e^{i(k_{x0}x+k_{y0}y)-i(\omega_{x,y}+\epsilon(k_{x0},k_{y0}))t}a^\dagger(b^\dagger)_{\{x,y\}}|0\rangle, \ee 
\end{widetext}
in order to create a beam with spatial Gaussian profile. In \eqref{source}, $w$ is the width of the beam, $\{x_0,y_0\}$ is the center of the source, $\{k_{x0},k_{y0}\}$ are the Bloch momentum of the beam, $\omega_{x,y}$ is the frequency of the resonator at coordinate $\{x,y\}$.

\begin{figure}
\centering\epsfig{file=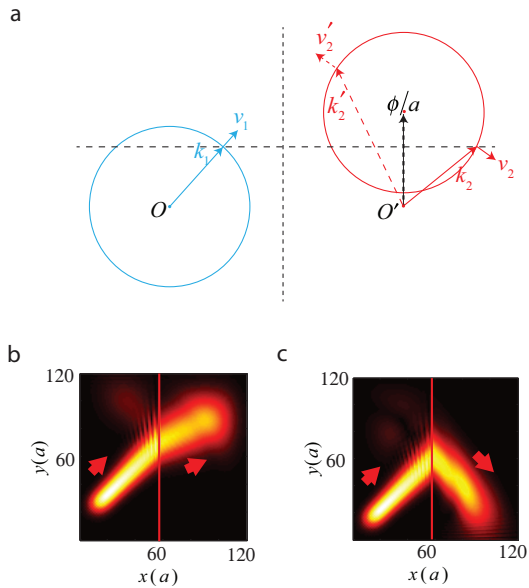,clip=1,width=0.8\linewidth,angle=0}\vspace{0.3cm}\hspace{0.3cm}
\caption{(Color online). \textbf{a} Analysis of beam refraction for the structure in Fig. \ref{lattice}, for $\phi(y)\equiv \phi$. Blue and red circles are the constant quasi-energy contours in the left and right regions, respectively. $k_{1,2}$ and $v_{1,2}$ are the incident and refracted momentum and group velocity of the beam. $k_1$ and $k_2$ have equal parallel component. $k_2^\prime$ and $v_2^\prime$ are the time-reversal of the refracted beam that corresponds to $k_2$. \textbf{b} Simulation for the case of $\phi=0.05$ showing positive refraction. \textbf{c} Simulation for the case of $\phi=2$ demonstrating negative refraction. }
\label{refraction}
\end{figure}

We first study the special case where in the right region $\phi(y)$ is a constant, i.e. $\phi(y)\equiv \phi$. In this case, both the left and the right regions have zero effective magnetic field. Nevertheless, we demonstrate that a beam propagating across the interface between these two regions undergoes refraction; and under certain condition, even negative refraction can happen. To understand such an effect, we investigate the band structures for the two regions of Fig. \ref{lattice}. For the left and right regions, the Floquet band structures are given by $\epsilon(k_x,k_y)=V[\textrm{cos}(ak_x)+\textrm{cos}(ak_y)]$ and $\epsilon(k_x,k_y)=V[\textrm{cos}(ak_x)+\textrm{cos}(ak_y-\phi)]$, respectively. Thus, in the momentum space, the band structure of the right region is shifted along the $y$ direction by $\Delta k_y=\phi/a$, as compared to the band structure of the left region. In general, arbitrary shift of the band structure in momentum space can be accomplished by appropriate choice of phase distribution. Such a capability for achieving a shift of the photonic band structure in momentum space is quite unique in our dynamically modulated systems, and has not been noted in any other system before. 

As a beam propagates through the interface between the left and right regions, the relation between the incident angle $\theta_i$ and refraction angle $\theta_r$ of a beam can be obtained by considering the conservation of quasi-energy and surface-parallel momentum. Suppose the momentum of the beam is small and thus the band structure in the left region can be approximated by $\epsilon(\vec k)\approx 2V-V (ak)^2$. The constant quasi-energy contour is a circle centered around $\vec k = 0$. At the same quasi-energy, the constant quasi-energy contour in the right region is shifted by $\phi/a$ along the $y-$axis as shown in Fig. \ref{refraction}a. Therefore, the relation between $\theta_i$ and $\theta_r$ becomes 
\be\label{nref} \textrm{sin}\theta_i-\textrm{sin}\theta_r=\frac{\phi}{a|\vec k_0|}, \ee
where $\vec k_0$ is the Bloch momentum of the incident beam \cite{note2}. From Eq. \eqref{nref}, we see that $\theta_r <0$, and hence negative refraction occurs, when $\phi/(2a)<|\vec k_0|$ and ${\rm max}\{{\rm sin}^{-1}[\phi/(a|\vec k_0|)-1],0\}<\theta_i<{\rm min}\{{\rm sin}^{-1}[\phi/(a|\vec k_0|)],\pi/2\}$. 

We verify the above analytical theory with direct numerical calculation using Eq. \eqref{sch}.  We take $\theta_i=45^\circ$, $|\vec k_0|=1/a$. In Fig. \ref{refraction}b, we assume $\phi = 0.05$ in the right region. In this case, the shift in the momentum space for the constant quasi-energy contour is small, and the beam passes through the interface with a small angle of refraction. In Fig. \ref{refraction}c, we assume $\phi = 2$, which generates a larger shift of the constant quasi-energy contour in the right region, and thus the beam undergoes negative refraction as it passes through the interface. Also, in both cases of Figs. \ref{refraction}b and c, we observe almost no reflection at the interface. This is due to the impedance matching between the two regions, since aside from the phases of the coupling constants all other parameters of the Hamiltonian are the same on both sides. 

The refraction of beam in this lattice may seem counter-intuitive, since the left and right regions have zero effective magnetic fields. Scrutinizing the structure of Fig. \ref{lattice} reveals that there are non-vanishing effective magnetic fields located at the interface of the two regions, since the phase accumulation around the plaquettes on the interface is non-zero. Thus, as an alternative to the previous explanation using the shift of band structure, one can equivalently state that the magnetic fields at the interface supply a canonical momentum (not the conserved Bloch momentum $k_y$) kick to the incident beam, leading to refraction. Such kind of magnetic flux induced beam refraction is difficult to observe for electrons, since one needs to achieve a magnetic field sheet with density of the magnetic flux quantum.

\begin{figure}[H]
\centering\epsfig{file=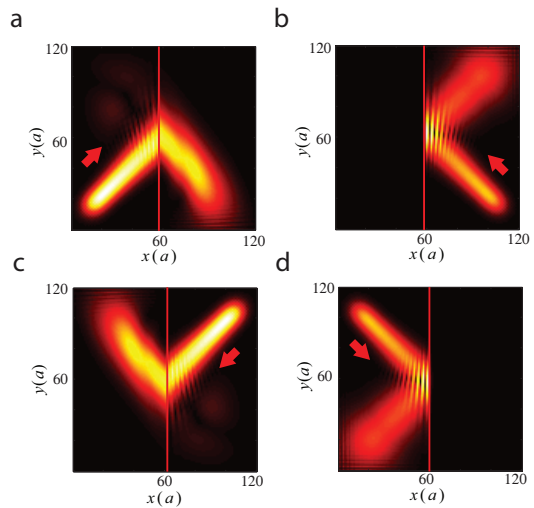,clip=1,width=0.8\linewidth,angle=0}\vspace{0.3cm}\hspace{0.3cm}
\caption{(Color online). Demonstration of the interface of the two regions in Fig. \ref{lattice} functioning as a circulator. The right region has $\phi = 2$. Red arrows indicate the incident direction of the beams.}
\label{mirror}
\end{figure}

In our structure, the momentum kick from the effective magnetic field at the interface is reminiscent of the concepts of meta-surfaces \cite{metas1, metas2}, or the nonlinear processes that give rise to negative refraction \cite{graphene}. However, unlike the meta-surfaces in Refs. \cite{metas1, metas2}, here the effective magnetic field breaks time-reversal symmetry. As a result, the beam propagation is non-reciprocal. As an illustration, we start with Fig. \ref{mirror}a, which reproduces Fig. \ref{refraction}b where a beam incident from the left region undergoes negative refraction as it passes through the interface. In Fig. \ref{mirror}b, we excite a beam, incident upon the interface from the right region, propagating along the direction opposite to the outgoing beam in Fig. \ref{mirror}a.  We observe instead total internal reflection at the interface. Such a total internal reflection can be accounted for by also examining Fig. \ref{refraction}a, where we note that in this case the incident beam from the right region (dashed arrows in Fig. \ref{refraction}a) can not excite any mode in the left region by momentum conservation consideration. 
Such a one-way total internal reflection has been previously considered in magneto-optical photonic crystals \cite{yu2007}. Here we achieve similar effect without the use of magneto-optics. 

Based on such one-way total internal reflection effect, the interface between the left and right regions in fact function as a four-port circulator, where the ports corresponding to the four incident beam directions as shown in the four panels of Fig. \ref{mirror}. Circulators has been previously considered for guided mode using either magneto-optical effect \cite{circulatorbook, wang2005} or with the use of dynamic modulation \cite{yu2009}. Here we show that a single interface can behave as a circulator for beams without guiding structure. 


The previous case corresponds to a constant $\phi(y)$ in the right region of the lattice. Next we consider the case where $\phi(y)$ is not a constant, which provides additional capabilities for beam manipulation. As a specific example, we design $\phi(y)$ to realize both on-axis and off-axis focusing effects for a collimated beam propagating along $x-$axis in the left region and normally incident onto the interface. We assume that the center of the incident beam is located at $y = 0$. To design $\phi(y)$ we follow a ray tracing procedure. For a ray incident upon a position $y$ at the interface, we choose $\phi(y)$ according to Eq. \eqref{nref} to achieve an angle of refraction that is appropriate for focusing. To create a focal point located at $(f,d)$ to the right of the interface, the ray tracing procedure above results in 
\be\label{phiy}
\phi(y)=ak_0\frac{y-d}{\sqrt{(y-d)^2+f^2}}.
\ee

\begin{figure}[H]
\centering\epsfig{file=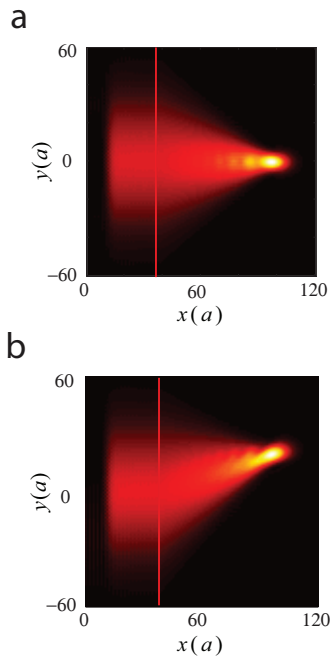,clip=1,width=0.5\linewidth,angle=0}\vspace{0.3cm}\hspace{0.3cm}
\caption{(Color online). A collimated beam is focused on-axis (a) and off-axis (b) by designing $\phi(y)$ according to Eq. \eqref{phiy} in the right region of Fig. \ref{lattice}. }
\label{lens}
\end{figure}

In Fig. \ref{lens}, using Eq. \eqref{phiy}, we generate two different modulation phase distributions. Fig. \ref{lens}a uses the parameters $f =60a$ and $d = 0a$, and Fig. \ref{lens}b uses the parameters $f = 60a$ and $d = 20a$. From the ray tracing procedure above, we expect on-axis focusing with respect to the beam axis in Fig. \ref{lens}a, and off-axis focusing in Fig. \ref{lens}b, which is indeed what we observe in the numerical simulations of Fig. \ref{lens}. Again, we emphasize that the two cases in Fig. \ref{lens} represent exactly the same structure, except with different modulation phase distributions, indicating versatile reconfigurability that is inherent in the use of such effective gauge field. 


Given the agreement between our analytic descriptions of the effect of the gauge fields, and the numerical simulations in two dimensions, below we will use the analytic description to design 3D structures. We extend the lattice of Fig. \ref{lattice} in $z$ direction to make a cubic lattice. The two kind of resonators are alternatively distributed in the cubic lattice with dynamically-modulated nearest neighbor coupling. We separate the lattice into two regions. In the left region ($x<0$), all the modulation phases on the bonds along the three axes are zero; in the right region ($x>0$), the phases on the bonds along the $x-$axis are all zero, but the phases along the $y-$axis ($\phi_y(y,z)$) and along the $z-$axis ($\phi_z(y,z)$) are functions of $y$ and $z$. To achieve a position-dependent refraction, the beam refraction across the interface at $x=0$ can be determined by the following equations:
\bqa
\label{2d1}\phi_y(y,z)=ak_0(\textrm{sin}\theta_{iy}-\textrm{sin}\theta_{ry}),\\
\label{2d2}\phi_z(y,z)=ak_0(\textrm{sin}\theta_{iz}-\textrm{sin}\theta_{rz}),
\eqa
where $\theta_{i(r)y,z}$ are the directional angle of incident (refracted) beam along $y-$ and $z-$axes respectively. Thus, the desired functionality, such as focusing, which is described by the $\theta^\prime$s, can then be implemented in our lattice through Eqs. \eqref{2d1}-\eqref{2d2}.


As final remarks, the experimental implementation of the Hamiltonian of Eq. \eqref{Ht} has been considered in Ref. \cite{our2}, and the key arguments are reproduced in the Supplementary Information. The dynamically modulated coupling between the two resonances, which is the crucial aspect that enables the creation of an effective gauge field, can be achieved using a mixer in the microwave frequency range, or a modulator in the optical frequency range. In the microwave frequency range, using such a mixer to create a gauge potential has already been demonstrated experimentally in Ref. \cite{our3}. In the experiment of Ref. \cite{our3}, the modulation phase is the phase of the local oscillator, which can be arbitrarily set after the structure is constructed. In the optical frequency range, the weak modulation strength ($\Delta n/n$) leads to a relative weak coupling between resonators; however the modulation phase again can be arbitrarily set by external modulation sources, as demonstrated in a recent experiment integrating silicon modulators \cite{tzuang}, and thus the amplitude of gauge field is not limited. Such a dynamic aspect of the gauge field, as well as the non-reciprocity generated by such a gauge field, differs significantly from several recent proposals and experiments that create an effective gauge field based on a spin degree of freedom in photons \cite{hafezi,umu,photonicTI,floquetphotonTI}. We also note the effects shown in the paper are robust to certain amount of resonant frequency disorders for practical considerations (see Supplementary Information). In conclusion, our work indicates significant new capabilities for controlling the spatial flow of light, through the control of temporal degrees of freedoms that generate an effective gauge field for photons. 

This work is supported in part by U. S. Air Force Office of Scientific Research grant No. FA9550-09-1-0704, and U. S. National Science Foundation grant No. ECCS-1201914.

\newpage

\setcounter{figure}{0}
\setcounter{equation}{0}

\makeatletter
\renewcommand{\thefigure}{S\@arabic\c@figure}
\renewcommand{\theequation}{S\@arabic\c@equation}
\makeatother
\renewcommand{\bibnumfmt}[1]{[S#1]}
\renewcommand{\citenumfont}[1]{S#1}

\section{Experimental implementation}
\begin{figure}[H]
\centering\epsfig{file=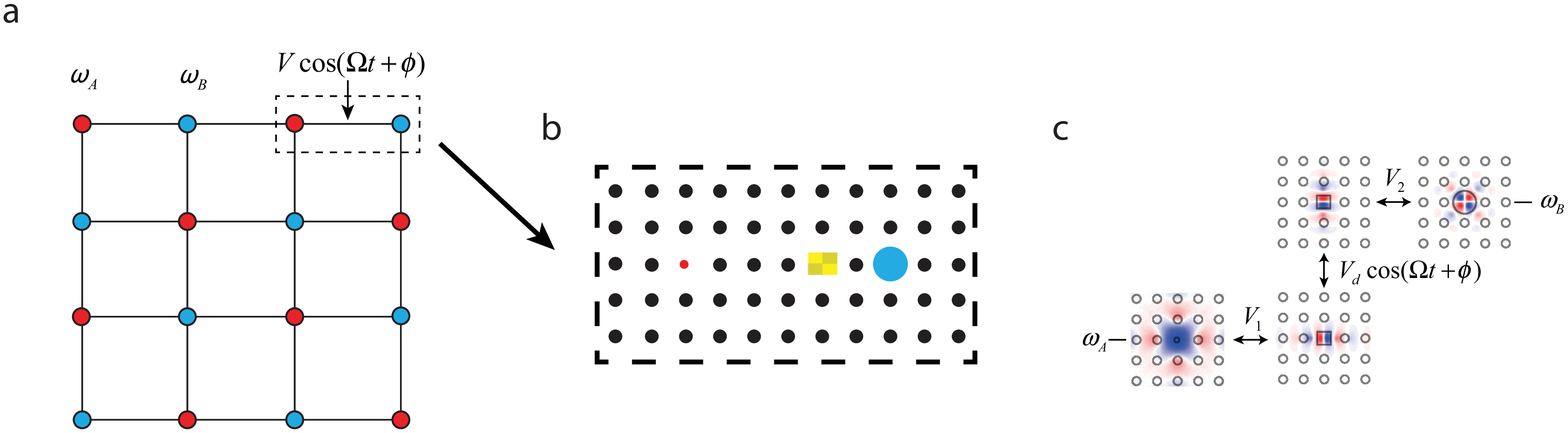,clip=1,width=1\linewidth,angle=0}\vspace{0.3cm}\hspace{0.3cm}
\caption{ \textbf{a} A photonic resonator lattice with harmonically modulated nearest-neighbor coupling. The red and blue dots correspond to resonators $A$ and $B$, respectively.   \textbf{b} A photonic crystal platform for implementing the dynamical coupling on a single bond between two resonators (dashed box in \textbf{a}). All dots here represent dielectric rods. The black dots here represent rods that create a photonic crystal. The periodicity here is approximately 540 nm for light with free space wavelength of 1.55 $\mu$m. The red and blue dots are rods with modified radii, and correspond to resonators $A$ and $B$, respectively. Resonator $A$ supports a monopole mode. Resonator $B$ supports a quadrupole mode. The yellow region is a coupling resonator supporting a pair of dipole modes. Its dielectric constant is modulated as a function of time. The details of the parameters can be found in Ref. \cite{our2}.  \textbf{c} Level diagram illustrating the field distributions, frequencies, and coupling, for the resonances in \textbf{b}.   }
\label{sifig}
\end{figure}

We discuss the physical implementation of the Hamiltonian of Eq. (1). This Hamiltonian describes a resonator lattice as schematically shown in Fig. \ref{sifig}(a), where the coupling constants between nearest-neighbor resonators are modulated dynamically in the form of $V\textrm{cos}(\Omega t+\phi)$. Conceptually, to implement this Hamiltonian, the key is to achieve such dynamical coupling between two resonators that form a single bond (dashed box in Fig. \ref{sifig}(a)). 

Here we introduce a physical structure based on point-defect resonators in a photonic crystal, as shown in Fig. \ref{sifig}(b), that provides such dynamic coupling. The structure in Fig. \ref{sifig}(b) contains resonators $A$ and $B$, supporting a monopole mode at frequency $\omega_A$, and a quadrupole mode at frequency $\omega_B$, respectively, as shown in Fig. \ref{sifig}(c).  Due to the frequency and symmetry mismatch between these two modes, they do not couple statically. To introduce a dynamic coupling between them, we place an additional coupling resonator between resonators $A$ and $B$. The coupling resonator supports a pair of dipole modes, at frequencies $\omega_A$ and $\omega_B$ respectively (Fig. \ref{sifig}(c)). The dielectric constant of the coupling resonator is modulated harmonically at a frequency $\Omega = \omega_B - \omega_A$, with a form
\be
\Delta\epsilon(\vec r, t) = \delta(\vec r)\textrm{cos}(\Omega t+\phi). 
\ee

The structure in Fig. \ref{sifig}(b) is therefore described by a four-mode coupled mode theory,
\bqa 
\frac{da_A}{dt}=i\omega_A a_A+iV_1 a_{p_x},\\ \frac{da_{p_x}}{dt}=i\omega_A a_{p_x}+iV_1 a_A+iV_d\textrm{cos}(\Omega t+\phi)a_{p_y},\\  \frac{da_{p_y}}{dt}=i\omega_B a_{p_y}+iV_2 a_B+iV_d\textrm{cos}(\Omega t+\phi)a_{p_x},\\ \frac{da_B}{dt}=i\omega_B a_B+iV_2 a_{p_y}, 
\eqa 
where $a_A$ and $a_B$ are the amplitudes in resonators $A$ and $B$ respectively, $a_{px}$ and $a_{py}$ are the amplitudes of the two modes in the coupling resonator. Under the condition $V_d\ll V_1, V_2, \Omega$, these equations can be further reduced to a two-mode coupled-mode theory equations, 
\bqa 
\label{twomode1}\frac{d\tilde a_A}{dt}=i\omega_A \tilde a_A+i\frac{V_d}{2}\textrm{cos}(\Omega t+\phi)\tilde a_B,\\\label{twomode2} \frac{d\tilde a_B}{dt}=i\omega_B \tilde a_B+i\frac{V_d}{2}\textrm{cos}(\Omega t+\phi)\tilde a_A,
\eqa
where $ \tilde a_{A(B)}=e^{iV_{1(2)}t}a_{A(B)}$ (Supplementary Information in Ref. \cite{our2}). The description of the effect of modulations on such a four-mode resonator system, in terms of the two-mode coupled mode theory, has been validated by direct finite-difference time-domain simulations as discussed in details in Ref. \cite{our2}.

We therefore see that the dielectric constant modulation of the coupling resonator generates a dynamic coupling of the required form $V\textrm{cos}(\Omega t+\phi)$ between the two single-mode resonators. Here the dynamic coupling constant is related to the dielectric constant modulation strength via
\be\label{vd}
V=\frac{V_d}{2}=\frac{\sqrt{\omega_A\omega_B}}{8}\frac{\int d\vec r\delta(\vec r)\vec E_1^\star\cdot\vec E_2}{\sqrt{\int d\vec r\epsilon(\vec r)|\vec E_1|^2}\sqrt{\int d\vec r\epsilon(\vec r)|\vec E_2|^2}},
\ee 
where $\vec E_{1,2}$ are the normal electric fields of the two dipole modes. And the phase $\phi$ is the same as the phase of the dielectric constant modulation. One should not confuse the phase $\phi$ of the dielectric constant modulation, which physically corresponds to the phase of the RF wave that generates the modulation, with the phase that the optical wave acquire as it propagates through the medium. We note that the gauge field arises from the spatial distribution of the modulation phase $\phi$, which can be arbitrarily set.

In general, the achievable dielectric constant modulation is weak, leading to a limited magnitude of $V$, which in turn places a constraint of the quality factor $Q$ of the resonators. In order that the photon amplitude does not diminish significantly after the beam steering, we require $V>\omega_{A,B}/Q$. This requirement is sufficient for the beam focusing effect shown in Fig. 4. For a modest modulation $\Delta\epsilon/\epsilon\approx 5\times10^{-5}$, we find $Q\approx 10^5$, which is achievable using the state-of-the-art photonic crystal resonator arrays \cite{resonator}.

\section{Effect of disorder in resonant frequencies}
We consider the effect of disorders that might be induced due to the modulation of the resonant cavities. The frequency of resonators can shift in the presence of dynamic modulation, given by a similar formula as Eq. \eqref{vd}:
\be\label{fshift}
\delta\omega=\frac{\omega}{8}\frac{\int d\vec r\delta(\vec r)\vec E^\star\cdot\vec E}{\int d\vec r\epsilon(\vec r)|\vec E|^2},
\ee 
where $\vec E$ is the electric field of the resonance mode. The frequency shift can be in principle eliminated by adopting an odd-symmetric profile $\delta(\vec r)$ of the modulation, as of the case of Fig. \ref{sifig}b, since thus the numerator of Eq. \eqref{fshift} will be zero. In the general case, where the symmetry of the modulation is not perfect, the modulation will introduce additional frequency fluctuation in the resonators. Here, we consider the general case where such kind of frequency shift is present, and show the beam propagation effects demonstrated in the paper are robust to reasonable amount of resonant frequency fluctuation as induced by dynamic modulation.

To simulate the effect of such a fluctuation in resonator frequency, we add a frequency shift term to each resonator in the lattice of Fig. \ref{sifig}a, $\delta\omega\textrm{cos}(\Omega t+\phi_r)$, where $\delta\omega$ is given by Eq. \eqref{fshift}, $\Omega$ is the modulation frequency, and $\phi_r$ is a phase taking a random value between 0 and 2$\pi$, and is generated for each resonator individually. Thus, the dynamically modulated lattice is dressed with time-dependent frequency disorders. The Hamiltonian (Eq. (1)) of the lattice is modified to be 
\begin{widetext}
\be\label{disorderHt} H(t)=\sum\limits_i \big(\omega_A+\delta\omega\textrm{cos}(\Omega t+\phi_{ri})\big)a^\dagger_ia_i+\sum\limits_j \big(\omega_B+\delta\omega\textrm{cos}(\Omega t+\phi_{rj})\big)b^\dagger_jb_j+\sum\limits_{\langle ij \rangle}V\textrm{cos}(\Omega t+\phi_{ij})(a^\dagger_ib_j+b^\dagger_ja_i).\ee
\end{widetext}
The Hamiltonian describes a resonator system where the instantaneous resonance frequencies fluctuate in both space and time. 

\begin{figure}[H]
\centering\epsfig{file=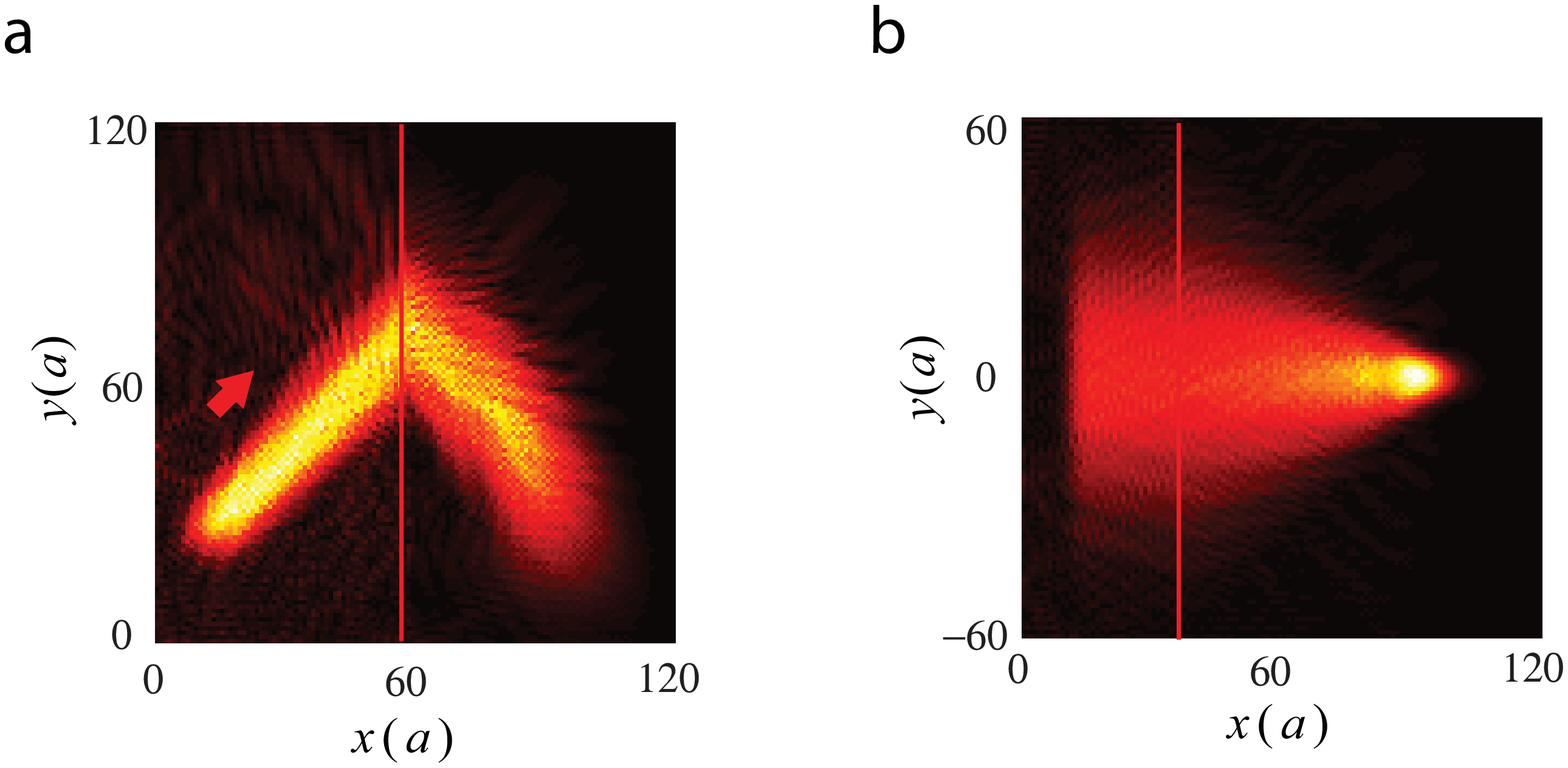,clip=1,width=1\linewidth,angle=0}\vspace{0.3cm}\hspace{0.3cm}
\caption{ Beam trajectories in the presence of disorders of resonant frequency: \textbf{a} refraction \textbf{b} focusing.  }
\label{sifig2}
\end{figure}

Using Eq. (2) and the Hamiltonian of Eq. \eqref{disorderHt}, Fig. \ref{sifig2} a and b show the beam trajectories corresponding to the refraction (Fig. 2c) and focusing (Fig. 4a) effects in the presence of the frequency disorders respectively, with $\delta\omega=V$. The major features of the trajectory are clearly preserved for such a disorder. Such robustness arises from the fact that the non-reciprocal phase induced by the modulation phase persists as long as the rotating wave approximation holds. A similar result also shows the effects are robust to static frequency disorders due to device fabrication.

\begin{references}
\bibitem{note2}
For larger beam momentum, Eq. \eqref{nref} should be modified using the exact constant quasi-energy contour determined by $\textrm{cos}(ak_x)+\textrm{cos}(ak_y)=\textrm{const.}$. However, similar beam control can still be achieved.
\end{references}
\end{document}